\documentclass{epl}

\newcommand{\C}{\textbf{C}}

\title{Spectral scalability as a result of geometrical self-similarity in fractal multilayers}
\author{S. V. Zhukovsky \and A. V. Lavrinenko \and S. V. Gaponenko}
\institute{
 Institute of Molecular and Atomic Physics, National Academy of Sciences of Belarus,
           Minsk 220072 Belarus, zhukovsky@imaph.bas-net.by}

\begin{document}
\maketitle
\begin{abstract}
The optical spectra of fractal multilayer dielectric structures
have been shown to possess spectral scalability, which has been
found to be directly related to the structure's spatial
(geometrical) self-similarity. Phase and amplitude scaling
relations, as well as effects of finite structure size, have been
derived.
\end{abstract}
\section{Introduction}
It is commonly known that the optical spectra of periodic
dielectric materials (including periodic multilayers in
particular) possess forbidden gaps, which are demonstrated to
directly result from spatial periodicity
\cite{Yariv,Yeh,Joannopoulos}. On the other hand, disordered
dielectric media have been discovered to slow down, localize, and
confine light waves traveling through them
\cite{Anderson,Lagendijk}. The same effects are known for
electrons and other quantum particles in periodic and random
potential, respectively (see \cite{Kronnig} and the review
\cite{Russell}). So, both periodic and random structures (which
represent the two extreme, and hence most studied, cases of
multilayers), exhibit characteristic spectral effects that result
from their topology.

Recent studies reveal that one type within the ``intermediate''
case (nonperiodic but deterministic structures) also displays
characteristic spectral effects not present in either of extreme
cases. It was found \cite{Kohmoto} that {\em quasiperiodic} (e.g.,
Fibonacci) multilayers have {\em self-similar} spectra. Their
transmission bands represent Cantor sets, a well-known example of
one-dimensional (1D) fractals. It was proved that spectral
self-similarity is a characteristic property of spatial
quasiperiodicity. The same is equally applicable for electronic
spectra.

In this Letter, we would like to address another class of
deterministic nonperiodic structures, namely {\em fractal
multilayers}. We show that their geometrical self-similarity
results in {\em spectral scalability}, earlier observed by us in
numerical computations (see Ref. \cite{PhysRevE}). This paper
analytically shows that the origin of scalability is the
self-similarity inherent to all fractal multilayers, and this is a
manifestation of correlation between geometrical properties of the
structures and properties of their eigenvalue spectra.


\section{Fractal multilayers}
One of common examples of fractal multilayers is a well-known
triadic Cantor stack generated using the ``middle third removal''
procedure \cite{Feder} (see Fig.\ref{fig1}a). However, this
procedure can be generalized. The most straightforward way to do
so is to complicate the removal routine, applying it not only to
the middle third, but to arbitrary (yet similar from generation to
generation) regions of the structure. Some variations are
described in \cite{Feder} and investigated in
\cite{PhysRevE,Jaggard1,Jaggard2,Sibilia}.

Here we introduce a more general procedure, which encompasses most
of 1D fractals that have prefractals and hence can be used in
multilayer design. The algorithm starts with an {\em initiator}, a
single dielectric layer (label it $A$) with refractive index $n_A$
and thickness $d_A$. The initiator is stacked together $G$ times,
and the layers are numbered in base $G$ (starting with zero).
Then, those parts whose numbers belong to a given subset of digits
$\C \subset \left\{0,1, \ldots ,G-1\right\}$ are replaced with
layers of another dielectric (labeled $B$), with refractive index
$n_B \neq n_A$ and thickness $d_B$. This replication-replacement
(RR) procedure is then repeated for the resulting structure (which
now consists of $G$ layers), with the only difference that a group
of $G$ $B$-type layers is now used to replace the appropriate
fragments. Repeating this RR procedure multiple times yields the
desired fractal multilayer.

Here, an arbitrary integer $G >2$ together with the subset $\C$
form the {\em generator} of the structure, while the number $N$ of
RR procedures applied is called the {\em number of generations}.
The whole structure can be referred to as a $(G, \C, N)$
structure. One can see that the usual $N$-stage middle third
Cantor stack is nothing but a particular case of $(3,
\left\{1\right\},N)$. Other particular cases include higher-$G$
Cantor structures ($G=3,5,7,\ldots$;
$\C=\left\{1,3,\ldots,G-2\right\}$,$N$) \cite{PhysRevE},
non-symmetric stacks \cite{Feder}, and generalized Cantor bars
\cite{Jaggard2}.

Sample stacks are shown in Fig.\ref{fig1}, and the construction
details can be inferred therefrom.

To conclude this section, let us list some simple but important
relations concerning fractal multilayers. First of all, the total
number of layers in such a structure is $G^N$ (here and further,
several adjacent layers of the same material count as separate
layers). Among these layers, $(G-C)^N$ are $A$-type and the rest
are $B$-type, $C$ being the number of members in $\C$. Then, the
total thickness of a $(G, \C, N)$ structure can be written as a
recurrent relation
\begin{equation}
\label{eq_thick}
\Delta_N = \left( {G - C} \right)^N d_A  + \left(
{G^N  - \left( {G - C} \right)^N } \right)d_B
\equiv
\left(G-C \right)\Delta_{N-1} + C \tilde \Delta_{N-1}.
\end{equation}

From Eq. (\ref{eq_thick}) one can obtain a scaling relation for
$\tilde \Delta_{N}=G^N d_B$, and in all cases $\Delta_0=d_A$. For
all calculations, the constituent layers were chosen have equal
optical thickness, i.e.,
\begin{equation}
\label{eq_qwave} n_A d_A = n_B d_B = d^* \equiv \pi c/2 \omega_0.
\end{equation}
This condition causes the spectra to be periodic with respect to
frequency, the period equal to $2 \omega_0 $. This outcome is very
convenient, since it provides a natural way to normalize the
frequency scale introducing the {\em dimensionless frequency}
$\eta \equiv \omega/\omega_0$. It also allows only {\em one}
period of spectrum to be referred to as ``spectrum'', which is
what will be done hereafter.

\section{Spectral scalability}
Keeping this in mind, we can now present a simple definition of
spectral scalability as follows. {\em We have found that the {\bf
whole} spectrum of a $(G, \C, N)$ stack appears as a {\bf part} of
a $(G, \C, N+1)$ stack spectrum}. If we magnify a certain part of
the latter (the area centered on $2 \omega_0$, or on
$\eta=0,2,\ldots$, to be exact) by a factor of $G$, its shape will
coincide very well with that of the former spectrum. This property
was observed and reported by us earlier \cite{PhysRevE} for $(3,
\left\{1\right\},N)$ and $(5, \left\{1,3\right\},N)$ structures
(see Fig.~\ref{fig2}a-e). The same method used in the
calculations, subsequent research has revealed that this property
holds for any $G$ and $\C$ (see, e.g., Fig.~\ref{fig2}f,g). So
does the relation for the factor by which one has to magnify the
central part of the $(G, \C, N_1)$ stack spectrum for matching
with that of $(G, \C, N_2<N_1)$ stack. Termed the {\em scaling
factor} between $(G, \C, N_1)$ and $(G, \C, N_2)$ structures, it
equals
\begin{equation}
\label{eq_scf}
S=G^{N_{1} - N_{2}}.
\end{equation}

The fact that the scaling factor in (\ref{eq_scf}) exactly equals
the geometrical factor of self-similarity, which is clearly seen
from the construction procedure, alone hints at the idea that {\em
geometrical self-similarity of fractal multilayers and scalability
of their optical spectra are related}. However, such qualitative
speculations are clearly not enough to state that spectral
scalability is a direct result of geometrical self-similarity.

A more convincing proof of this statement may be obtained from
analytical calculations. It is worth noting, however, that
spectral scalability, while visually apparent as in
Fig.~\ref{fig2}, is difficult to be described mathematically
because close inspection of the spectra reveals that there is no
exact coincidence either in the value of transmission coefficient
or in the peak locations (see Fig.~\ref{fig3}a). However, these
discrepancies do not change the shape of the spectral curve
noticeably, thus not hindering the observation of scalability.

First we consider the simplest case, the middle third Cantor
stacks $(3, \left\{1\right\}, N)$. To analytically calculate the
spectra of such multilayers, it is possible to use the {\em
self-similarity method of calculation} \cite{Jaggard2,Jaggard1},
which is a generalization of Airy formulas based on the structure
being self-similar. According to this method, the reflection and
transmission coefficients for the $(3, \left\{1\right\}, N+1)$ and
$(3, \left\{1\right\}, N)$ structures are related as
\begin{equation}
\label{eq_sj_rt}
R_{N + 1} \left(\eta \right) = g_r \left[ {R_N (\eta ),T_N
(\eta ),\tilde \Delta _N,\eta } \right], \;
T_{N + 1} \left(\eta \right) = g_t \left[ {R_N (\eta ),T_N
(\eta ),\tilde \Delta _N,\eta } \right],
\end{equation}
where $\tilde\Delta_N$ is as defined by Eq. (\ref{eq_thick}). The
functions
\begin{equation}
\label{eq_sj_g}
g_r (x,y,d,\eta ) =
x + \frac{{xy^2 \varepsilon^2\left(d,\eta\right)}}
{{1 - x^2 \varepsilon^2\left(d,\eta\right)}},\;
g_t (x,y,d,\eta ) =
\frac{{y^2 \varepsilon\left(d,\eta\right)}}
{{1 - x^2 \varepsilon^2\left(d,\eta\right)}},
\end{equation}
are obtained using effective medium formalism in \cite{Jaggard1}.
The initial conditions for these recurrent relations are derived
from the normal-incidence reflection and transmission coefficients
for a single layer (i.e., a structure with $N=0$), such that
\begin{equation}
\label{eq_sj_zero}
R_0 \left(\eta \right) = -r +
\frac{rtt'\varepsilon_0^2\left(\Delta_0,\eta\right)}
{1 - r^2 \varepsilon_0^2\left(\Delta_0,\eta\right)},
\;
T_0 \left(\eta \right) =
\frac{{tt'
\varepsilon_0\left(\Delta_0,\eta\right)}}
{{1 - r^2 \varepsilon_0^2\left(\Delta_0,\eta\right)}}.
\end{equation}
Here $r$, $t$, $t'$ are normal-incidence Fresnel's coefficients
for the layer interfaces, and
\begin{equation}
\label{eq_eps}
\varepsilon\left(d,\eta\right) \equiv \exp \left[{{\textstyle{i \over c}} \eta \omega_0 n_B d} \right], \;
\varepsilon_0\left(d,\eta\right) \equiv \exp \left[{{\textstyle{i \over c}} \eta \omega_0 n_A d} \right]
\end{equation}
are the phase exponents.

Now, to proceed with the analysis of scalability, we need to
compare the following quantities
\begin{equation}
\label{eq_comp}
\begin{array}{ccc}
T_{N+1}\left(\textstyle{\eta \over 3}\right)
& \textrm{and} &
T_N\left(\eta \right).
\end{array}
\end{equation}

Using the relation for $\tilde\Delta_N$ and substituting
(\ref{eq_sj_rt}) and (\ref{eq_sj_g}) into (\ref{eq_comp}), one can
obtain
\begin{equation}
\label{eq_expand1}
T_{N+1}\left(\textstyle{\eta \over 3}\right)=
\frac{{T_N^2 \left( {\textstyle{\eta  \over 3}} \right)
\exp \left[ {{\textstyle{i \over c}}{\textstyle{\eta  \over 3}} \omega_0 n_B  \cdot 3^N d_B } \right]}}
{{1 - R_N^2 \left( {{\textstyle{\eta  \over 3}}} \right)
\exp \left[ {{\textstyle{{2i} \over c}}{\textstyle{\eta \over 3}}\omega_0 n_B \cdot 3^N d_B } \right]}},
\end{equation}
\begin{equation}
\label{eq_expand2}
T_N\left(\eta \right)=
\frac{{T_{N - 1}^2 \left( {\Delta _{N - 1} ,\eta } \right)
\exp \left[ {{\textstyle{i \over c}}\eta \omega_0 n_B \cdot 3^{N - 1} d_B } \right]}}
{{1 - R_{N - 1}^2 \left( {\Delta _{N - 1} ,\eta } \right)
\exp \left[ {{\textstyle{{2i} \over c}}\eta \omega_0 n_B \cdot 3^{N - 1} d_B } \right]}}.
\end{equation}

We see that the phase exponents in (\ref{eq_expand1}) and
(\ref{eq_expand2}) are exactly equal, and the sole difference
between $T_{N+1}\left(\textstyle{\eta \over 3}\right)$ and
$T_N\left(\eta \right)$ lies in the coefficients,
$T_{N}\left(\textstyle{\eta \over 3}\right)$,
$R_{N}\left(\textstyle{\eta \over 3}\right)$ and
$T_{N-1}\left(\eta \right)$, $R_{N-1}\left(\eta \right)$,
respectively. But if one expands these coefficients in the same
way, using (\ref{eq_sj_rt}) and (\ref{eq_sj_g}), one can see that
the difference will again manifest itself only in the
coefficients, this time, $T_{N-1}\left(\textstyle{\eta \over
3}\right)$, $R_{N-1}\left(\textstyle{\eta \over 3}\right)$ and
$T_{N-2}\left(\eta \right)$, $R_{N-2}\left(\eta \right)$,
respectively.

Tracing this procedure down to $N=0$ and seeing that all
frequency-dependent exponents that appear along the way are equal
for both terms in (\ref{eq_comp}), we finally reach the point
where subsequent substitution of Eqs. (\ref{eq_sj_rt}) is no
longer possible. At this point the factors to be compared are
$T_{1}\left(\textstyle{\eta \over 3}\right)$,
$R_{1}\left(\textstyle{\eta \over 3}\right)$ and $T_{0}\left(\eta
\right)$, $R_{0}\left(\eta \right)$. The corresponding phase terms
are $\varepsilon\left(\Delta_0,\eta\right)$ and
$\varepsilon_0\left(\Delta_0,\eta\right)$ as defined in
Eq.~(\ref{eq_eps}), and they are equal if the condition
(\ref{eq_qwave}) is met. The difference in coefficients is smaller
as $r$ decreases, and the agreement is total if $r^2 \approx 0$.

As we have seen, all frequency-dependent exponents in the
expressions (\ref{eq_comp}) are equal at any stage of
decomposition. So it can be said that the quantities in
Eq.~(\ref{eq_comp}) have {\em identical phase structure}, with a
minor difference in the coefficients. Since the characteristic
spectral features (transmission resonances and local band gaps)
are essentially phase phenomena (resulting from constructive and
destructive interference, respectively), similar phase structure
results in similar appearance of spectral portraits as confirmed
by Fig.~\ref{fig2}.

Rigorous analytical generalization of these results to all fractal
multilayers is rather straightforward and can be achieved by
further generalizing the Sun-Jaggard computation procedure, e.g.,
according to the multiple-reflection effective-medium formalism
presented in \cite{Antonio}.

\section{Amplitude mismatch: vertical scalability}
However, as can be seen in Fig.~\ref{fig3}a, in the areas {\em
between} characteristic features there is a significant difference
in transmittance value $T$. This difference, which does not alter
the shape of spectra, can be eliminated if, in addition to the
above mentioned frequency scaling (\ref{eq_scf}), the value of
$T_{N+1}\left(\eta/G\right)$ is raised to a certain power $\gamma$
(see Fig.~\ref{fig3}b). Thus, the final {\em scalability equation}
has the form
\begin{equation}
\label{eq_scale}
\left[T_{N+1}\left(\textstyle{\eta \over G}\right)\right]^\gamma = T_N\left(\eta\right).
\end{equation}

Numerical analysis reveals that $\gamma$ depends on the structure
parameters and varies slightly with frequency. In the region close
to $\eta=0$, where scalability is most often observed, we have
found $\gamma$ to equal
\begin{equation}
\label{eq_gamma}
\begin{array}{ccc}
\gamma = \alpha+\frac{1-\alpha}{f^2},& & f \equiv \frac{G-C}{G},
\end{array}
\end{equation}
where $\alpha$ was found to be small ($\alpha \cong 0.1$, see Fig.
\ref{fig4}).

It is important to note that $\gamma$ {\em only} depends on the
ratio between $G-C$ and $G$, called the {\em dielectric filling
fraction}, and so e.g., the structures $(6,
\left\{2,3\right\},N)$, $(6, \left\{3,4\right\},N)$, $(6,
\left\{1,4\right\},N)$ have the same $\gamma$. The fact that it
does not depend on the position of ``removed'' layers, nor on the
structure's lacunarity, nor on the value of $n_B/n_A$, together
with an observation that in the frequency region in question the
propagating wave exhibits little internal reflection, suggests
that {\em it is only the amount of dielectric ``removed'' during
the transition $N \to N+1$ that is important for $\gamma$}. Thus,
we can move forward to conclude that the only important parameter
in $f$ [see Eq.~(\ref{eq_gamma})] is its numerator $G-C$, while
the sole part of the denominator $G$ is to allow for the frequency
scaling (\ref{eq_scf}).

That said, it is enough to investigate the simplest case in order
to analytically establish the relation $\gamma(f)$. Instead of an
arbitrary $(G, \C, N)$, consider an equivalent structure $(G,
\C'=\left\{G-C,\ldots,G-1\right\}, N)$, which is effectively a
single layer whose thickness scales as
\begin{equation}
\label{eq_scd} d_N = \left(G-C\right)^N d_A.
\end{equation}

Using the Airy formulas for the transmission we can rewrite the
scalability equation
\begin{equation}
\label{eq_scmod1}
\left| {
\frac {\left( {1 - r^2 } \right)\varepsilon'}
{1 - r^2 \varepsilon'^2 }} \right|
= \left|
{\frac
{\left( {1 - r^2 } \right)\varepsilon'{}^f }
{1 - r^2 \varepsilon' {}^{2f} }} \right|^\gamma
\Rightarrow
\frac{1+r^4-2r^2\cos\delta}{1+r^4-2r^2} =
\left( \frac{ 1+r^4-2r^2\cos f\delta}{1+r^4-2r^2}\right)^\gamma
\end{equation}
where $\varepsilon' \equiv e^{i\delta}=\exp[\left(G-C\right)^N
\eta\pi/2]$.

Since we are staying close to $\eta=0$, we can assume $\eta \ll 1$
and therefore $\delta \ll 1$. At the 2{\it nd} order of Taylor
series of $\cos \delta$ (the first order obviously leads to an
identity $1=1^\gamma$),
we finally arrive at
\begin{equation}
\label{eq_scmod}
1+\frac{2r}{\left(1-r\right)^2} \delta^2 =
1+\gamma\frac{2r}{\left(1-r\right)^2}\left(f\delta\right)^2
 \Rightarrow
\gamma = \frac{1}{f^2},
\end{equation}
which has a good agreement with the numerically obtained
Eq.~(\ref{eq_gamma}) (see Fig.~\ref{fig4}).
An even better analytical agreement can be achieved using a finer
approximation for $\cos\delta$. This also results in a weak
dependence $\gamma(\eta)$, as was numerically confirmed and found
not significant for the observation of scalability.

\section{Peak mismatch: perturbation in characteristic effects}
So far we have shown that both phase structure matching and
amplitude matching can be derived analytically. However, in
Fig.~\ref{fig3} one can observe small mismatches in the resonance
peak locations for the spectral curves in (\ref{eq_scale}). This
agrees with the difference in the coefficients for the quantities
(\ref{eq_comp}) and shows that spectral scalability is only {\em
approximate} in real multilayers. However, we state that it
results from the {\em finite size} of the structures under study.
So, they are in fact prefractals rather than true fractals, so
spatial self-similarity in them is not exact either. This disturbs
the scalability effect in much the same way as it occurs in other
types of media. For example, finite periodic structures cannot
exhibit completely zero transmission in the band gaps, and in
finite disordered media light cannot be completely trapped. In
this manner, $N${\em th} generation Cantor multilayers can be
compared to $N$-period 1D photonic crystals, while it is commonly
known that band gaps are prominent at much larger $N$ than were
used for the plots in Fig.~\ref{fig3}.

However, if certain conditions are met, one can observe decent
band gaps even in periodic multilayers with as many as four
periods. An analogous statement is true for scalability in fractal
multilayers. But the condition to be desired is opposite. As was
noted earlier and confirmed in numerical calculations, the peak
mismatch decreases if the refractive index contract is small,
while band structure is more pronounced if the contrast is large
enough \cite{Yariv,Yeh}.

%
Had it been otherwise, i.e., if $N$ approached infinity, it is our
guess that spectral scalability would be exact. This can be
indirectly confirmed by plotting the scaled spectra for several
generations. As seen in Fig.~\ref{fig5}, the mismatch goes smaller
as $N$ grows larger.

\section{Conclusion}
To summarize, using the method that inherently contains spatial
self-similarity \cite{Jaggard1} along with scaling relations
(\ref{eq_scf}) and (\ref{eq_scd}), we have found that fractal
multilayers exhibit scalability both in phase and in the value of
transmittance according to the Eq.(\ref{eq_scale}). So, it can be
concluded that {\em spectral scalability is actually the result of
spatial self-similarity}, and moreover, it is a {\em
characteristic relation} between a topological property of a
multilayer structure and a spectral property of wave propagation.
These results are also applicable for the electronic spectra in a
fractal potential.

\acknowledgements The authors wish to acknowledge I.~S.~Nefedov,
who first introduced to us the ideas of making use of spatial
self-similarity in calculations, and to thank A.~G.~Smirnov for
critical reading of the manuscript. This work was supported by
INTAS (2001-0642) and in part by the Basic Research Foundation of
Belarus (grant no. F03M-097).


\begin{figure}
\caption{Sample fractal multilayer structures:
(a) Cantor ``middle third'' structures $(3, \left\{1\right\}, N)$
with the relations for $\Delta_N$ and $\tilde\Delta_N$ shown;
(b) higher-$G$ Cantor structures $(5, \left\{1,3\right\}, N)$;
(c) non-Cantor fractal structures $(4, \left\{1\right\}, N)$ for smaller values
of $N$.}
\label{fig1}
\end{figure}

\begin{figure}
\caption{Scalability of optical spectra for fractal multilayers:
(a) $(3, \left\{1\right\}, N=4)$, the central part of the spectrum
magnified in the frequency scale by 3 versus (b) the full period
of the $N=3$ spectrum; (c) the central part of $N=4$ magnified by
$9=3^2$, (d) the central part of $N=3$ magnified by 3, and (e) the
full period of the $N=2$-spectrum; (f) $(4, \left\{1\right\}, 4)$,
the central part of the spectrum magnified by 4 versus (g) full
period of the $(4, \left\{1\right\}, 3)$. Compare the looks of (a)
and (b); (c), (d), and (e); (f) and (g).}
\label{fig2}
\end{figure}

\begin{figure}
\caption{Exact comparison of the scaled spectra for $(3,
\left\{1\right\}, N)$ structures: (a) unchanged, (b) raised to a
power $\gamma$.}
\label{fig3}
\end{figure}

\begin{figure}
\caption{The dependence $\gamma\left(f\right)$. The dots are
numerical data, and the solid curve is the best-fit function. The
dashed line is the analytically derived function.}
\label{fig4}
\end{figure}

\begin{figure}
\caption{An enlargement of a small part of the
$(3, \left\{1\right\}, N)$ spectra. $N$ ranges from 3 to 8. The
peak mismatch is smaller as $N$ is larger.}
\label{fig5}
\end{figure}

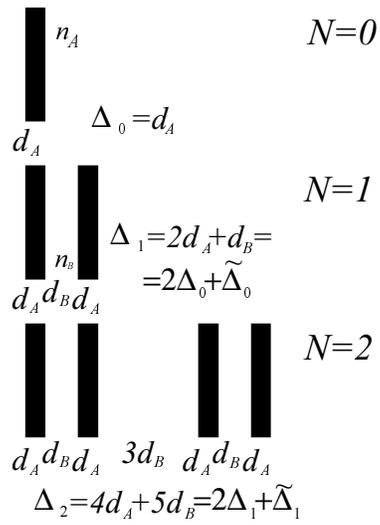
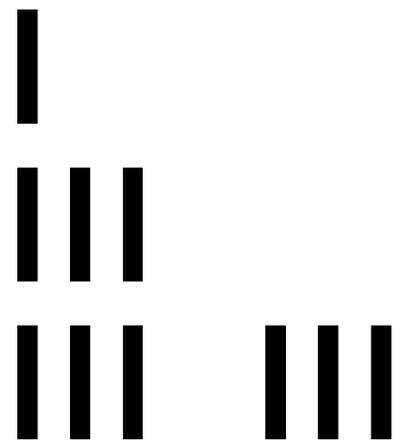
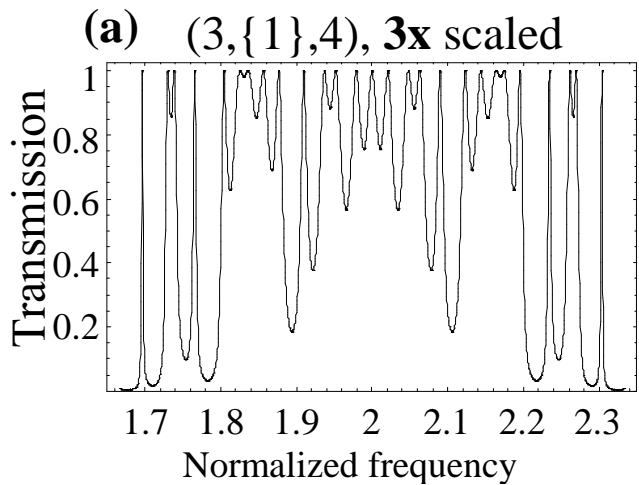
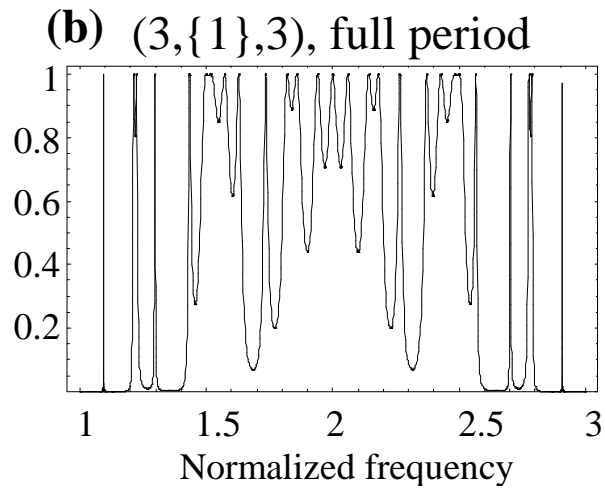
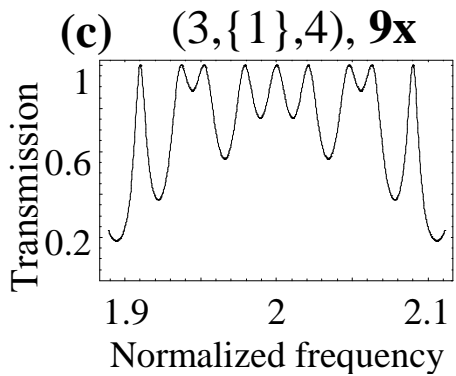
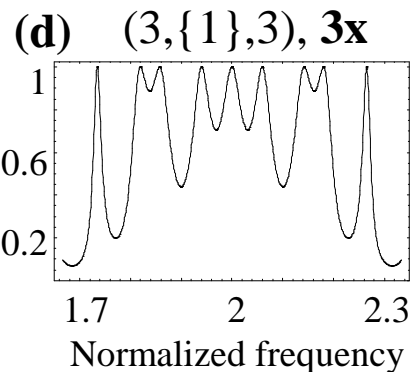
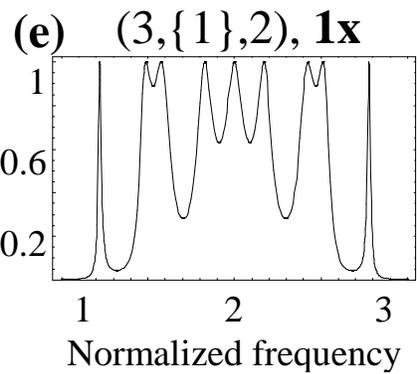
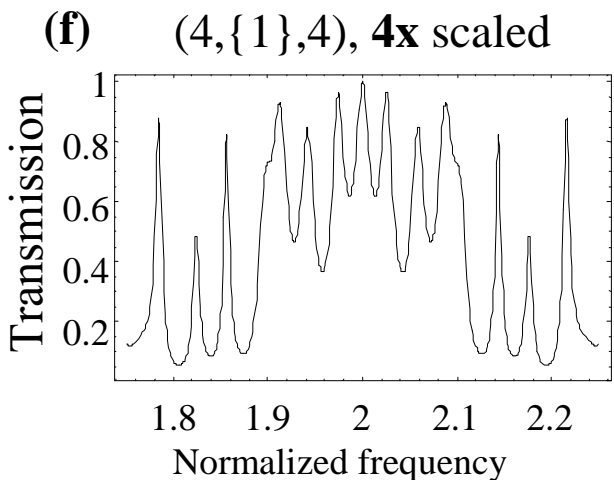
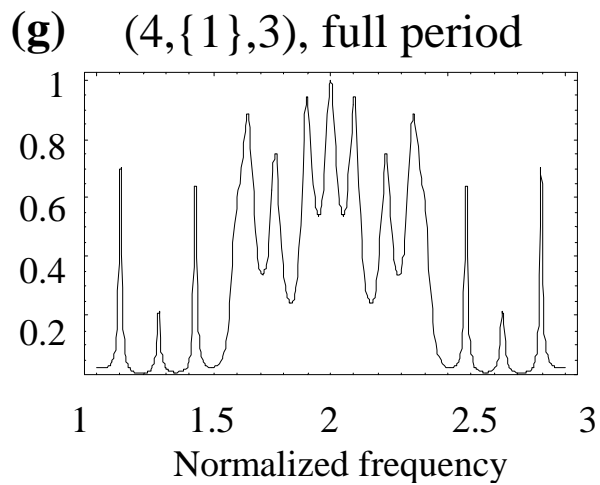
\end{document}